\documentclass[pra,
                preprint,
                amsmath,
                amssymb,
                showpacs,
                superscriptaddress]{revtex4}
\usepackage{graphicx} 
\usepackage{dcolumn}  
\usepackage{bm}       
\usepackage{amsfonts}
\usepackage{dsfont}
\usepackage{xcolor}

\begin{document}

\title{Simultaneous control of harmonic yield and energy cutoff of high-harmonic generation using seeded plasmonic-enhanced fields}

\author{Irfana N. Ansari}
\affiliation{%
Department of Physics, Indian Institute of Technology Bombay,
            Powai, Mumbai 400076, India}

\author{M. S. Mrudul}
\affiliation{%
Department of Physics, Indian Institute of Technology Bombay,
            Powai, Mumbai 400076, India}
            
\author{Marcelo F. Ciappina}
\affiliation{Institute of Physics of the ASCR, ELI-Beamlines project, Na Slovance 2, 18221 Prague, Czech Republic}

\author{Maciej Lewenstein}
\affiliation{ICFO - Institut de Ciencies Fotoniques, The Barcelona Institute of Science and Technology, Av. Carl Friedrich Gauss 3, 08860 Castelldefels (Barcelona), Spain}
\affiliation{%
ICREA, Pg.~Llu\'is Companys 23, 08010 Barcelona, Spain}

\author{Gopal Dixit}
\email[]{gdixit@phy.iitb.ac.in}
\affiliation{%
Department of Physics, Indian Institute of Technology Bombay,
            Powai, Mumbai 400076, India}
            
\date{\today}



\begin{abstract}
We study high-order harmonic generation (HHG) driven by seeded plasmonic-enhanced fields. On one hand, plasmonic-enhanced fields have shown a great potential to extend the HHG cutoff, an instrumental pre-requisite for the generation of attosecond pulses. On another hand, the use of XUV seeds appears to have a considerable potential to improve the HHG conversion efficiency, which is typically modest when a unique fundamental laser pulse is  employed. By mixing these two sources, we show it is possible to, simultaneously, boost the HHG cutoff and to increase the harmonic photon flux.  The combination of these features potentially enables to  generate intense and spectrally broad attosecond pulse trains. 
\end{abstract}

\maketitle 
\section{Introduction}
Higher-order harmonics are generated as an outcome of the interaction between a strong laser  and 
gaseous atoms, molecules and recently solid state materials. High-harmonic generation (HHG) is a highly non-linear process where 
an intense infrared radiation is up-converted into high frequency radiation in extreme ultraviolet (XUV) 
and soft x-ray frequency regime.  Not only the resulting radiation is used to synthesise bright and  coherent 
attosecond pulses~\cite{ferray1988multiple, corkum1993plasma, lewenstein1994theory, bartels2002generation, sansone2006isolated, paul2001observation, krausz2009attosecond} 
but, also used to probe multi-electron and
nuclear dynamics in atoms and molecules on their natural timescale~\cite{haessler2010attosecond, 
baker2006probing, smirnova2009high, li2008time, frumker2012oriented, dixit2012, bredtmann2014x, 
lepine2014attosecond, sansone2010electron, gruson2016attosecond}. 

The microscopic mechanism of HHG is well-understood by appealing to the so-called three-step model: 
As a first step, an electron is tunnel ionised due to modification of the Coulomb potential  by the strong laser field. 
In second step, the ionised electron is driven back and forth by the laser electric field and finally, 
the  electron  recombines with the parent ion and emits high-harmonic radiation~\cite{corkum1993plasma, lewenstein1994theory}. The maximum photon energy emitted in HHG is given by the cutoff law $E_{\textrm{max}} = I_{p}+3.17 U_{p}$, where $I_{p}$ is the ionisation potential of  the target atom or molecule and $U_{p}$ is the 
cycle-averaged ponderomotive energy of the free-electron. This energy can be increased either by choosing a target 
atom with higher $I_{p}$ value or by increasing $U_{p}$, which depends on the frequency $\omega$ and intensity $I$ of the driving laser pulse as $U_{p} = I/4 \omega^{2}$. 
The ionisation saturation of atoms and molecules puts an upper limit to the intensity of the laser pulse that can be used. Also, the efficiency of harmonic generation falls drastically as the wavelength is increased~\cite{tate2007scaling, frolov2008wavelength}. 
A better scheme to  increase the cutoff energy is based on the plasmonic field enhancement method,
where the surface plasmon polaritons developed in nano-structures are exploited to enhance the local electric field.  
Plasmonic field enhanced HHG offers an invaluable tool to probe attosecond electron dynamics at nanoscale and near-field spectroscopy of nanostructures~\cite{kruger2018attosecond, kruger2012attosecond, ciappina2017attosecond}.

In the pioneering work, Kim {\it et al.} have demonstrated that the driving laser field is enhanced  by 
several orders of magnitude in the vicinity of a bowtie shaped nanostructure. 
Under this scheme, high  intensity XUV radiation with wavelengths down to 47 nm in argon 
has been generated~\cite{kim2008high}. 
Besides the controversies~\cite{sivisletter,kimreply}, the experiment of  Kim {\it et al.} have motivated a truly deluge of experimental and theoretical investigations on HHG driven by 
plasmonic field enhancement in structured nano-objects~\cite{park2011plasmonic, pfullmann2014nano, han2016high, stebbings2011generation, sivis2013extreme, pfullmann2013bow, kruger2018attosecond, kruger2012attosecond, ciappina2017attosecond}. 

The amount of intensity enhancement crucially depends on the shape of the nanostructure. 
Various shaped nanostructures,  like 
nanoparticles~\cite{yang2013high}, metal nanotips~\cite{kruger2018attosecond, kruger2012attosecond}, 
metallic waveguides~\cite{park2011plasmonic} and plasmonic antennas~\cite{sivis2013extreme}, have been used to obtain  plasmonic field enhancement.  Husakou {\it et al.} has provided one of the first detailed theoretical investigations of HHG driven by a field generated in the vicinity of metal nanostructures~\cite{husakou2011theory}. Later on, a series of theoretical and numerical works have been carried out to understand and explain the underlying  mechanisms of plasmonic-enhanced HHG in more general 
scenarios~\cite{ciappina2012high, ciappina2012enhancement, ciappina2014high, ciappina2014coherent, shaaran2012estimating, shaaran2013quantum, shaaran2012quantum, 
shaaran2013high, perez2013beyond, ebadian2017extending, ebadi2014interferences, feng2015attosecond, feng2013attosecond, fetic2013high, he2013wavelength, wang2014high, wang2013control, husakou2011polarization, luo2014efficient, chacon2015numerical, he2013wavelength, neyra2018high, yavuz2012generation, yavuz2013gas}. 

Another instrumental aspect of HHG is the conversion efficiency, i.e.~the ratio of the measured harmonic beam energy to the driving laser energy. The control in the efficiency of the harmonic generation can be achieved by controlling macroscopic parameters such as focus position and pressure of the gas jet  of atoms or molecules, amongst others.  
On the other hand, manipulation and control of the frequency-time characteristics of HHG at microscopic level 
has been explored in depth. In this regard, various combinations of driving laser
fields along with seed pulses (attosecond pulse trains or a single attosecond pulse) have been used to enhance the generation efficiency of high-order harmonics, e.g.,~a dramatic enhancement 
of HHG has been reported in Ref.~\cite{bandrauk2002attosecond}, quantum path control in Refs.~\cite{schafer2004strong, gaarde2005large}, and attosecond control of ionization in Refs.~\cite{ishikawa2007single, ishikawa2004efficient, takahashi2007dramatic}.
A vertical electronic transition is induced by the additional seed pulse, which reduces the tunnelling barrier significantly and enhances the ionisation rate of tunnelling. As a result of the tunnel rate boost, the harmonic yield enhances by several orders of magnitude, in comparison to the harmonics generated by the fundamental driving field alone.

In this present work, we demonstrate  the possibility of simultaneous enhancement of the harmonic yield and increase in the HHG cutoff energy. For this purpose, high-order harmonics are generated by combining a fundamental plasmonic-enhanced driving  pulse and an XUV seed pulse. The frequency of the XUV seed pulse is in resonance with the first excited electronic state of the target atom. In a sense, our contribution follows up a series of two-colour plasmonic-enhanced studies. For instance, Cao {\it et al.} have shown the procedure of generating isolated sub-10 attosecond pulse by using spatially inhomogeneous two-colour $\omega-2\omega$, laser pulses~\cite{cao2014generation}. Furthermore, ultraviolet-assisted mid-infrared plasmonic fields are used to generate a supercontinuum efficiently~\cite{luo2014efficient}.  Our numerical analysis is performed by solving the time-dependent Schr{\"o}dinger equation (TDSE) in reduced dimensions. A time-frequency map of the HHG, obtained by Gabor transformation, is used to get a detailed insight  about the mechanisms of simultaneous enhancement of the HHG yield and the increase in the HHG cutoff energy. 

The organisation of the paper is as follows: Sec. II presents the theoretical model and numerical method to simulate HHG driven by  seeded plasmonic-enhanced fields; whereas results and discussion are presented in Sec. III. The conclusion of the paper is given in Sec. IV. Atomic units are used throughout in this paper unless specified otherwise. 

\section{Theoretical Method} 
The electron dynamics, induced by the strong laser field, is mostly limited along the polarisation direction of the laser electric field.  
In this work, the polarisation is considered to be linear along the $x$-axis.  
To model the interaction of our model atom with the external plasmonic-enhanced laser field, we solve the one-dimensional (1D) TDSE in length-gauge:
\begin{equation}\label{eq01}
i\frac{\partial\psi(x,t)}{\partial t}= \left[-\frac{\partial^{2}}{2 \partial x^{2}} + V_{\textit{atom}}(x) -x E(x, t) \right] \psi(x,t),
\end{equation}
where $V_{\textit{atom}}(x) = -\frac{1}{\sqrt{x^{2}+a}}$ is a soft-Coulomb potential to model the target atom. 
For $a = 1.0$,  the ionisation potential of the model atom is found to be 0.67 a.u. (18.2 eV)~\cite{ciappina2012high}.    

\begin{figure}[]
\includegraphics[width= 17cm]{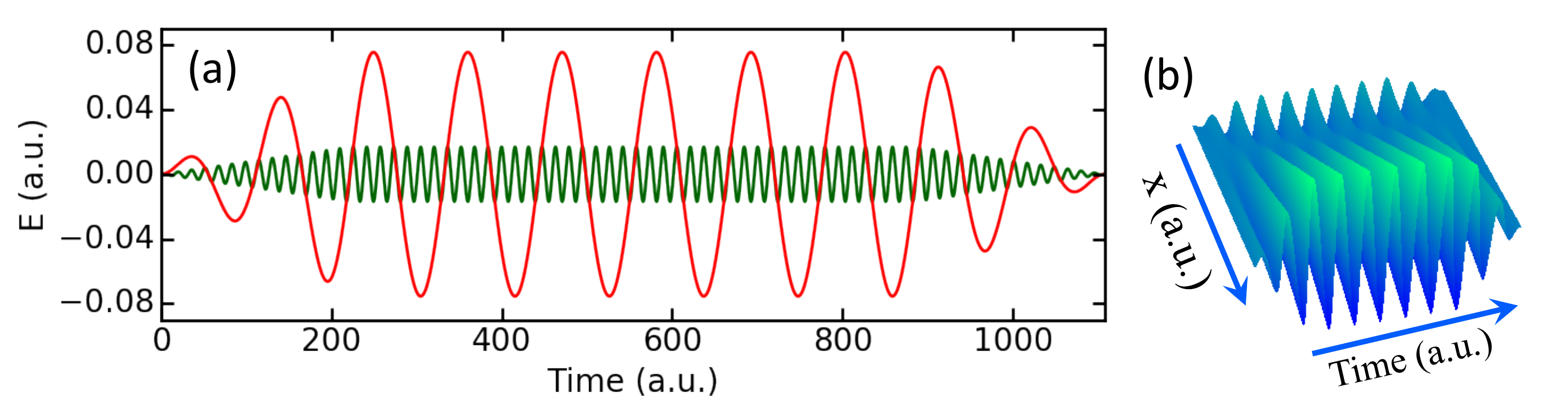}
\caption{(a). The form of the homogeneous electric fields of the fundamental (red) and seed (green) pulses. (b) The space-dependent inhomogeneous  electric field of the fundamental pulse. } \label{fig01}
\end{figure}

The total plasmonic-enhanced electric field becomes space-dependent at a nanometric scale and substantially modifies the electron dynamics~\cite{husakou2011theory, ciappina2012high, yavuz2012generation}. The space-dependent inhomogeneous electric field can be written in a general form as:
\begin{equation}\label{eq02}
E(x, t) = (1+\beta f(x)) E(t), 
\end{equation}
where $\beta$ is a parameter used to characterise  the strength of inhomogeneity. In this work, we have considered the simplest form of the function $f(x)$, i.e., $f(x)=x$. The decision to choose this linear form is motivated by previous works on HHG mediated by plasmonic-enhanced fields~\cite{husakou2011theory, ciappina2012high, yavuz2012generation, cao2014generation}. In this case, $\beta$  has the dimension of inverse length and $\beta = 0$ mimics the situation of spatially homogeneous fields.
The total time-dependent electric field, consisting of the fundamental and seed laser pulses, has the following form: 
\begin{equation}\label{eq03}
E(t) = g(t)[E_{0} \sin(\omega t) + E_{\textrm{seed}} \sin (n \omega t)], 
\end{equation}
where $g(t)$ is the pulse envelope.  Both the pulses have the same trapezoidal envelop with two optical cycles turned-on and turned-off and a plateau of six optical cycles of the fundamental frequency $\omega$. The total duration of the pulse is approximately 27 fs as shown in Fig.~\ref{fig01}.  The fundamental driving pulse has an intensity of $1\times10^{14}$ W/cm$^{2}$, whereas the seed pulse has a weaker value: $1\times10^{13}$ W/cm$^{2}$.  The fundamental pulse has a frequency equal to $\omega = 0.0567$ a.u. (corresponding to a wavelength $\lambda=800$ nm and a photon energy of 1.55 eV)  and its seventh-harmonic, i.e., $n = 7$ (corresponding to a $\lambda_{\textrm{seed}} \simeq 115$ nm and a photon energy of 10.8 eV) is chosen as the seed frequency. 
The seed pulse frequency is selected in close resonance with the energy of the  first excitation of the model atom.

To obtain the eigen-states and eigen-values  of the 1D-Time-independent Hamiltonian,  diagonalisation scheme is used. 
The split-operator method is used to propagate the electron wave-packet and solve the 1D-TDSE~\cite{hermann1988split}. 
The total wavefunction is multiplied by a mask function of the form $\cos^{1/8}$ to avoid spurious reflections from the spatial grid boundaries. The mask function  varies from 1 to 0 starting from the 2/3 of the spatial grid~\cite{ciappina2012high}.  The time-dependent dipole is calculated in the acceleration gauge as 
\begin{equation}\label{eq04}
a(t) = -\langle \psi (t) \vert \nabla_{x} V (x,t)\vert \psi (t) \rangle, 
\end{equation}
where $V (x,t) = V_{\textit{atom}}(x) -x E(x, t)$. Finally, the high-order harmonic spectrum is computed as the modulus squared of the Fourier transform of Eq.~(\ref{eq04}).

\section{Results and Discussion}    

 \begin{figure}[]
\includegraphics[width= 17cm]{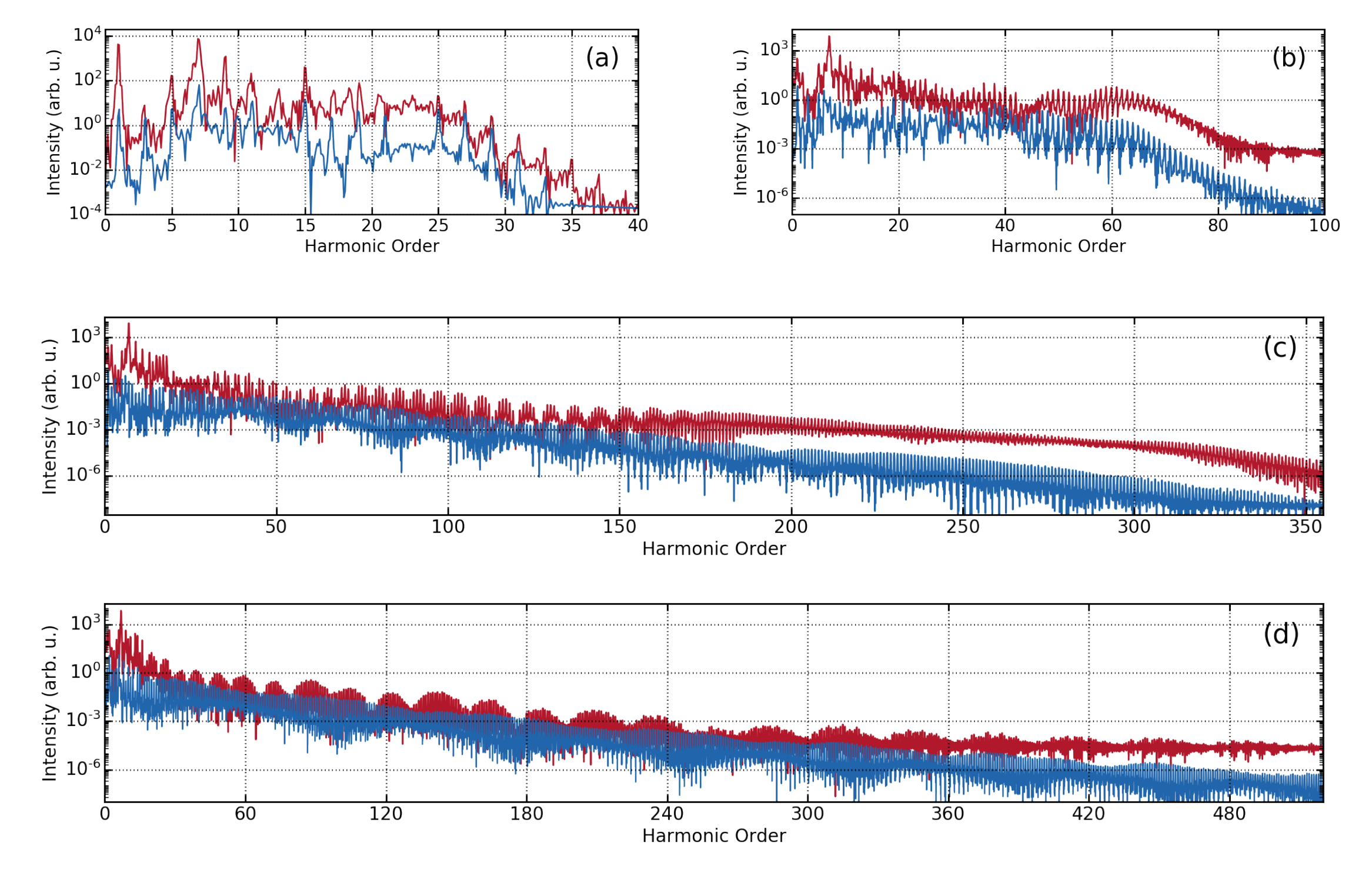}
\caption{
High-order harmonic spectrum for a model atom, with ionization potential $I_p=0.67$ a.u., obtained by solving TDSE in one-dimension for a spatial grid of $x_{\textrm{lim}} = \pm 7.5 \alpha_{0}$, where  
$\alpha_{0}$ is the quiver radius (see the text for details). The spectrum  generated by only driving fundamental laser  field 
is shown in blue colour, whereas the spectrum obtained by combined driving fundamental and seed 
laser  fields is shown in red colour.  The harmonic spectra for (a) homogeneous laser field $(\beta = 0$); for inhomogeneous field strength parameter (b) $\beta = 0.01$, (c) $\beta = 0.02$, and (d) $\beta = 0.05$ are shown. The  inhomogeneous field strength parameter $\beta$ has a dimension of length inverse, so $\beta = 0.01, 0.02~\textrm{and}~0.05$ correspond to inhomogeneity regions of 100 a.u. (5.3 nm), 50 a.u. (2.7 nm), and 20 a.u. (1 nm), respectively.} \label{fig1}
\end{figure}

Figure~\ref{fig1} presents HHG spectra for various situations. The spectra shown in blue and red colours are 
generated by the fundamental laser field only and by the combination of fundamental and seed laser fields, respectively. 
A below-threshold seventh-harmonic ($n = 7$) is used as a seed, which is in resonance with the first excited state of our model atom (the first electronic excited state has an energy of $-0.28$ a.u.,i.e.,~$-7.49$ eV).  
The extension of the spatial grid $x_{\textrm{lim}}$ is considered as an additional variable in solving the 1D-TDSE and mimics the situation where the electron is moving in a confined region~\cite{ciappina2012high}. 
The spatial region where the electron dynamics takes place is expressed in terms of the quiver radius $\alpha_{0} = E_{0}/\omega^{2}$ and the value of $\alpha_{0} \simeq 23$ a.u. (1.2 nm) corresponds to a fundamental driving field of 
800 nm wavelength with an intensity of $1\times10^{14}$ W/cm$^{2}$.  
The results presented in Fig.~\ref{fig1} are simulated for $x_{\textrm{lim}} = \pm 7.5  \alpha_{0}$, which is equal to a gap of 18.7 nm between the apexes in a bowtie shaped nanostructure~\cite{kim2008high, ciappina2012high}.    

The HHG spectra for various inhomogeneous field strengths are shown in Fig.~\ref{fig1}(a-d). The harmonic spectrum for an homogeneous field, ($\beta = 0$) is shown in Fig.~\ref{fig1}(a), whereas the spectra for inhomogeneous fields  corresponding to $\beta = 0.01, 0.02~\textrm{and}~0.05$ are, respectively, presented in Figs.~\ref{fig1}(b), \ref{fig1}(c) and \ref{fig1}(d).  
As evident from the spectra, we observe a drastic boost of the HHG energy cut-off as the strength of the inhomogeneous field increases. 
As usual, the intensity of the harmonics is decreasing gradually with the harmonic order, developing an abrupt cutoff for the cases of Figs.~\ref{fig1}(a)-(b). On the contrary, there is no  sharp energy cutoff in the case of the higher 
inhomogeneous field strengths, Figs.~\ref{fig1}(c)-(d). 
This behaviour is related with the modifications in the electron trajectories, which occur as a consequence of the spatial dependence of the laser electric field (see below for more details). The drastic increase in HHG energy cutoff could be understood by considering the spatial nature of plasmonic-enhanced electric field. As the electron moves away from the parent ion, the field strength experienced by the electron increases, resulting in an increment of its velocity and hence the photon energy. 

One more interesting observation can be seen from Fig.~\ref{fig1}: the spectra exhibit several minima and show an oscillatory behaviour in the intensity profile. The strength of these oscillation is more pronounced when the harmonics are generated using the combination of fundamental and seed pulses in the presence of the inhomogeneity. For $\beta = 0.01$, the intensity of spectrum is slowly varying and shows few minima only. On the other hand, ~\ref{fig1}(c) ($\beta = 0.02$) exhibit rapid oscillations for seeded harmonics (red) and several minima are present till 170th harmonic order. As the inhomogeneity strength increases to $\beta = 0.05$, the frequency  of the minima in the intensity profile decreases and the presence of the minima is extended to higher photon energies (Figs.~\ref{fig1}(c) and ~\ref{fig1}(d)). This means that, the overall oscillation profile does not seem to have any direct relationship with the strength of the inhomogeneity.
With the addition of the spatial inhomogeneity, the even order harmonics are also present. This is expected as the symmetry of the system is broken by the introduction of the space-dependent electric field. Furthermore, with the introduction of the seed pulse, the efficiency of the HHG yield is enhanced  by few-orders of magnitude (red colour)  in comparison  to the harmonics generated  by the fundamental pulse only (blue colour).  
In order to gain a deeper insight about the mechanism responsible for the increment of the HHG energy cutoff and intensity modulation, we perform a Gabor analysis of the HHG spectra.  

The Gabor transformation of the time-dependent dipole is performed as:
\begin{equation}\label{eq05}
a_{G}(\Omega, t) = \int dt^{\prime}  \frac{\textrm{exp}[-(t-t^{\prime})^{2}/2 \sigma^{2}]}{\sigma \sqrt{2 \pi}} \textrm{exp}(i \Omega t^{\prime}) a(t^{\prime}).
\end{equation} 
To get an adequate balance between the time and energy resolutions,  we have considered $\sigma = 1/(3 \omega)$   in the present work. 
By performing the Gabor transformation, a time-frequency map of the HHG spectra is obtained. 

\begin{figure}[h!]
\includegraphics[width=17cm]{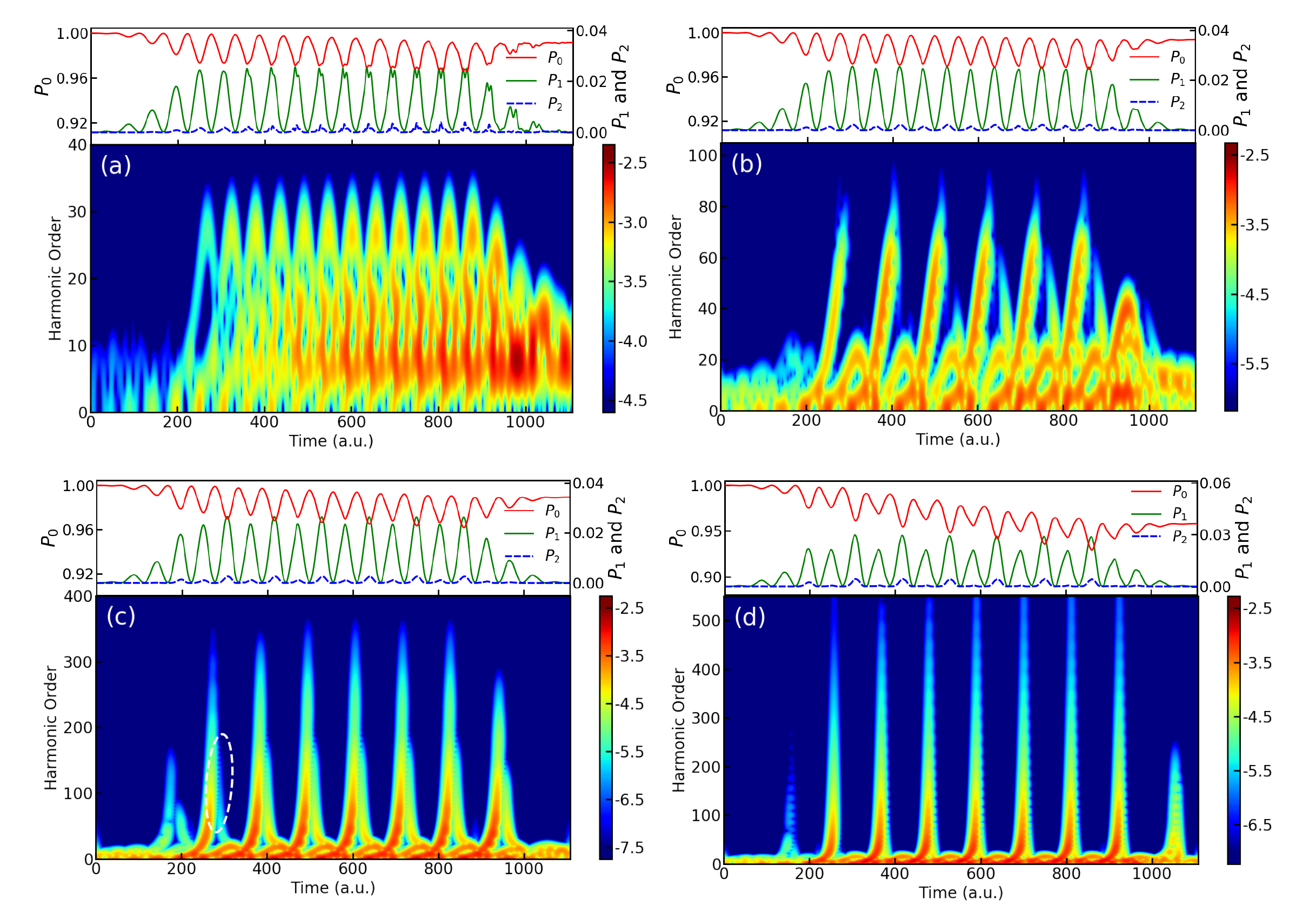}
\caption{Populations of the ground ($P_{0}$, in red), first ($P_{1}$, in green) and second ($P_{2}$, in blue) excited  states; and  time-frequency map of the HHG spectra 
generated by the fundamental laser field only (shown by blue colour in Fig.~\ref{fig1}).  
(a) homogeneous laser field $(\beta = 0$); (b) $\beta = 0.01$, (c) $\beta = 0.02$, and (d) $\beta = 0.05$.} \label{fig2}
\end{figure}

Figure~\ref{fig2} represents
time-frequency maps of the HHG spectra obtained for the fundamental laser field only (shown in Fig.~\ref{fig1}, blue colour),  joint with the populations of the ground and excited states (see below for more details).  
A pair of short and long trajectories is generated in each half-cycle of the laser electric field for the spatial homogeneous case (Fig.~\ref{fig2}(a)). With the introduction of the space-dependent electric field, the contribution from one pair of trajectories is enhanced whereas the other pair is unaffected within one laser cycle (Fig.~\ref{fig2}(b)). Also, a clear enhancement in the short trajectories contribution is observed for the higher-order harmonics. On the contrary, the long trajectories are suppressed for such cases. Furthermore, the weight of the long trajectory increases with each cycle of the electric field. 
 
For higher strengths of the spatial inhomogeneity,  the electrons developing long trajectories have higher kinetic energies and contributes till approximately the 200th to 500th harmonic orders for $\beta = 0.02$ and 0.05, respectively (Figs.~\ref{fig2}(c) and ~\ref{fig2}(d)).  Also, the recombination of the electron with the parent ion occurs much faster after the change of the direction of the driving electric field. This leads to a temporal overlap between the short and long trajectories and gives a clear interference pattern as shown by the highlighted region (white ellipse in Fig.~\ref{fig2}(c)). The oscillation frequency is large till the 100th harmonic order for  $\beta = 0.02$, and later it continuously decays with harmonic orders. 
As evident from Fig.~\ref{fig2}(c), 
the profile of this interference pattern clearly matches with the oscillation profile of the HHG spectrum as shown in Fig.~\ref{fig1}(c). Also, the oscillation frequency is smaller for $\beta = 0.05$ in comparison to $\beta = 0.02$, and this is in accordance with the harmonic spectrum presented in Fig.~\ref{fig1}(d). Hence, this interference effect completely explains the oscillatory behaviour of the harmonic spectra.

At the top of each time-frequency map in Fig.~\ref{fig2}, the populations of the electronic states are also presented. The ground state ($P_{0}$), first ($P_{1}$) and second ($P_{2}$) excited states populations are shown in red, green and blue colours, respectively.
As evident from the figure, 
the populations of the ground ($P_{0}$) and first excited states ($P_{1}$) change in every half cycle, whereas the populations of the other excited states do not change significantly. 
As the strength of the fundamental driving field reaches its maximum, 
the population of $P_{1}$ shows 
a double peak structure. The interference of the two-pair of trajectories is a reason behind this double peak structure. As the character of the driving field changes from homogeneous to inhomogeneous, the contribution  of  one pair of  trajectory is suppressed. This results 
in the absence of the double peak structure in the population $P_{1}$ (green). 
As the plasmonic field strength increases with the  inhomogeneity parameter, the population $P_{1}$ remains unchanged, whereas the depletion in the ground state population increases significantly from approximately 1.5 $\%$ for $\beta = 0.02$ to  5 $\%$ for $\beta = 0.05$ (Figs.~\ref{fig2}(c) and \ref{fig2}(d), red colour). 
This behavior in the wavepacket dynamics could be related with the modifications predicted in H$_+^2$ molecules when are
driven by plasmonic enhanced fields~\cite{yavuz2016,yavuz2018}. 
\begin{figure}[h!]
\includegraphics[width=17cm]{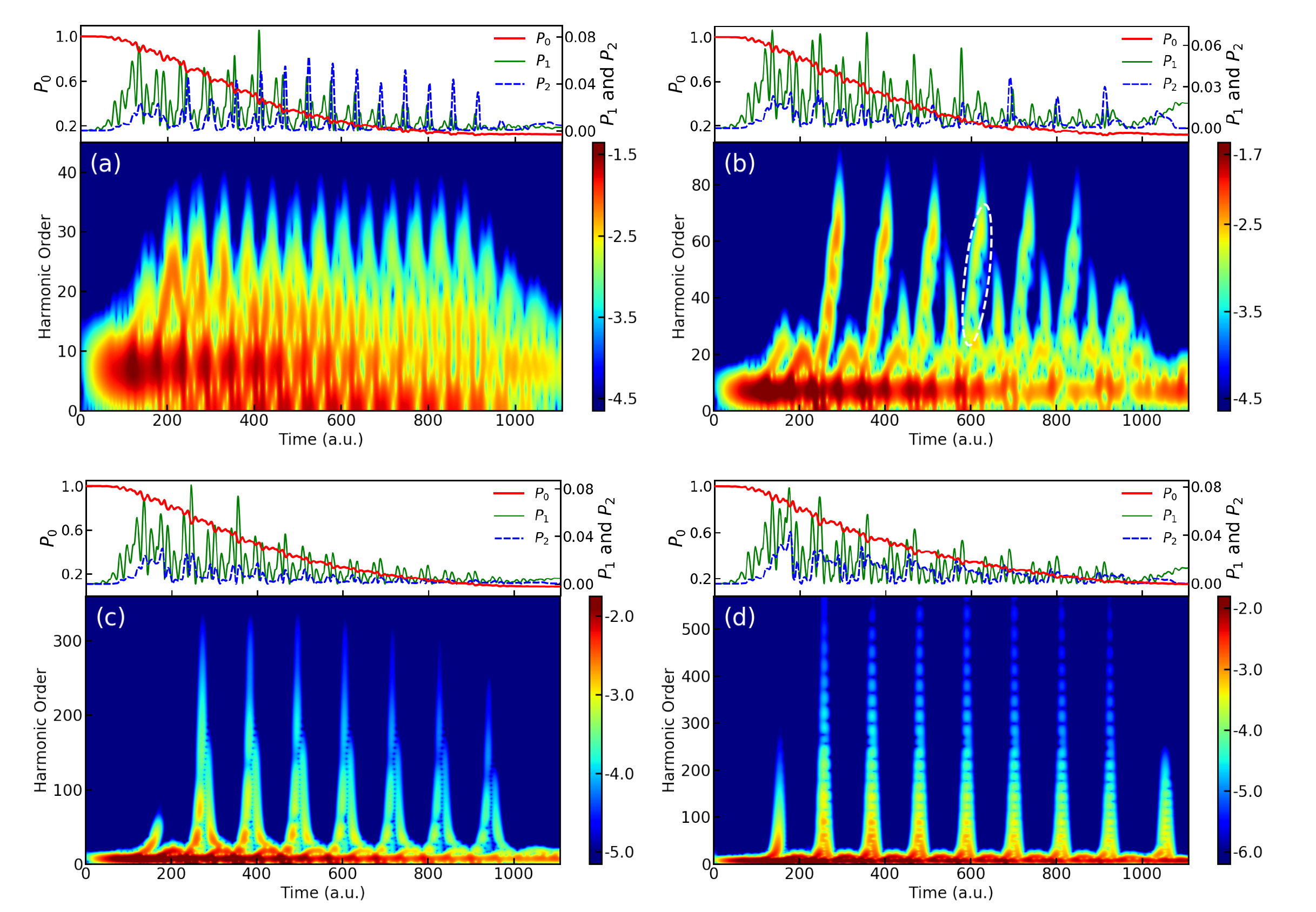}
\caption{ Populations of the ground ($P_{0}$, in red), first ($P_{1}$, in green) and second ($P_{2}$, in blue) excited  states; and time-frequency map of the HHG spectra 
generated by the combination of the fundamental and seed laser fields (shown by red colour in Fig.~\ref{fig1}).  
(a) homogeneous laser field $(\beta = 0$); (b) $\beta = 0.01$, (c) $\beta = 0.02$, and (d) $\beta = 0.05$.} \label{fig3}
\end{figure}

When a seed pulse, resonant with the first excited state, is introduced along with the fundamental driving pulse, some ground-state population is transferred to
the first excited state. This population transfer  opens several possibilities for HHG, e.g. the electron can be ionised from the ground (first excited) state and 
recombine to the first excited (ground) state, or the electron can ionise from and recombine to the ground (first excited) state. 
These new possibilities for the HHG are manifested by the population dynamics of electronic states. 
Likewise Fig.~\ref{fig2}, the populations of the three states: ground state ($P_{0}$, in red), first ($P_{1}$, in green) and second ($P_{2}$, in blue) excited states are presented at the top of each time-frequency map in Fig.~\ref{fig3} for the seeded harmonics. 
The population of the ground state drops drastically in all the four situations, i.e., for $\beta = 0.00, 0.01, 0.02$ and 0.05. The population is depleted by more than 85 $\%$ at the end of the combined fundamental and seed pulses.
This situation is vastly different compared with the case when HHG is driven by the  fundamental laser field only, where
the maximum depletion in the ground state population was a merely $5\%$.
This drastic loss of the  ground state population explains the reduced strength of the Gabor profile during the later cycles of the pulse.
For homogeneous field ($\beta = 0.00$),  
the population dynamics of the excited states is completely different from the previous case (Fig~\ref{fig2}), e.g.~the population dynamics of the second excited state ($P_{2}$, blue colour) plays a significant role in the HHG, which was almost absent for
the case where only the fundamental pulse drove the HHG process. 
Moreover, there is a significant dynamics of the first excited state population ($P_{1}$, green colour) at the beginning of the pulse and slowly the populations dynamics of $P_{2}$ 
catches up.  As the inhomogeneity strength increases, the role of 
second excited state diminishes (Figs.~\ref{fig3}(c), and ~\ref{fig3}(d); blue colour).  Moreover, the precise periodicity of $P_{0}$ and $P_{1}$ is lost for the seeded harmonics in comparison to the previous case (Fig.~\ref{fig2}).

The time-frequency maps of the harmonic spectra, corresponding to Fig.~\ref{fig1}, generated by the combination of fundamental and
seed laser fields are shown in Fig.~\ref{fig3}. As evident from the panels, with the introduction of the seed pulse, the 
 intensity of the harmonics is enhanced by few-orders in magnitude.  Due to the presence of many pairs of trajectories generated by the seed pulse, as well as the fundamental one, it is difficult to differentiate the contribution from 
 different trajectories for harmonic orders less than twenty. For $\beta = 0.01$, the  time-frequency map shows an oscillatory behaviour in the intensity of the short trajectories with respect to  the frequency of the emitted photon (highlighted by a white ellipse in Fig.~\ref{fig3}(b)).  
The presence of this oscillation in the time-frequency map is overshadowed by the interference of electron trajectories for the higher values of inhomogeneity, i.e., $\beta = 0.02$ and 0.05 (Figs.~\ref{fig3}(c) and~\ref{fig3}(d)). Even though the interference effect is dominant, a change in the short trajectories intensities can be seen for $\beta = 0.02$. The positions of these minima are different in each short trajectory and are following the excited states population dynamics. When there is a significant population in the first excited state, the minima are at lower harmonic orders (energies), whereas when the population of the excited state decreases, these minima are located at larger photon energies. 
 In the presence of the seed pulse, the interference effect of long and short trajectories becomes much more pronounced for higher strengths of inhomogeneity.

 \begin{figure}[]
\includegraphics[width= 17cm]{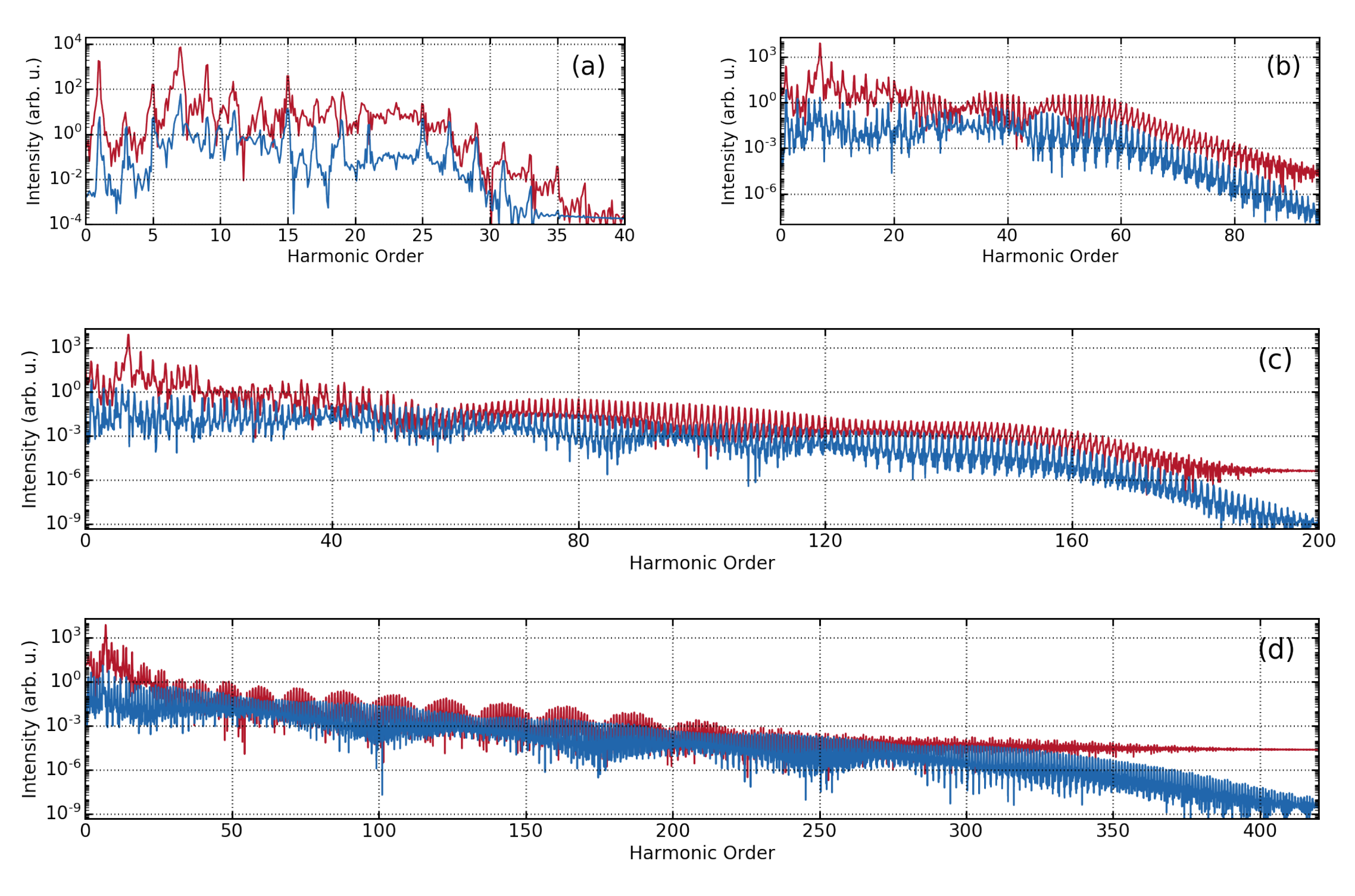}
\caption{(colour online). Same as figure~\ref{fig1} for $x_{\textrm{lim}} = \pm 4.5 \alpha_{0}$.} \label{fig4}
\end{figure}

An additional parameter used in the TDSE simulations is the extension of the spatial grid. 
The spatial region controls the electron dynamics and this is reflected in the characteristics of the HHG spectra . 
By reducing the size of the spatial grid, the spatial extent for the electron dynamics is reduced. 
Figure~\ref{fig4} presents the harmonic spectra for $x_{\textrm{lim}} = \pm 4.5 \alpha_{0}$ (a gap of 11.0 nm between the apexes)
keeping all the other parameters identical to  Fig.~\ref{fig1}. 
In comparison to the harmonic spectra obtained for $x_{\textrm{lim}} = \pm 7.5 \alpha_{0}$, 
the spectra for  $\beta = 0.00$ and 0.01 are similar whereas  the energy cut-off is reduced significantly for 
$\beta = 0.02$ and 0.05.  The oscillation in the intensity of the harmonics fades out as the 
motion of the electron is confined  by reducing $x_{\textrm{lim}}$. 
However, the efficiency of the seeded harmonics are higher by few-orders of magnitude in comparison to the harmonics generated by the fundamental pulse only. 
 
\begin{figure}[h]
\includegraphics[width= 17cm]{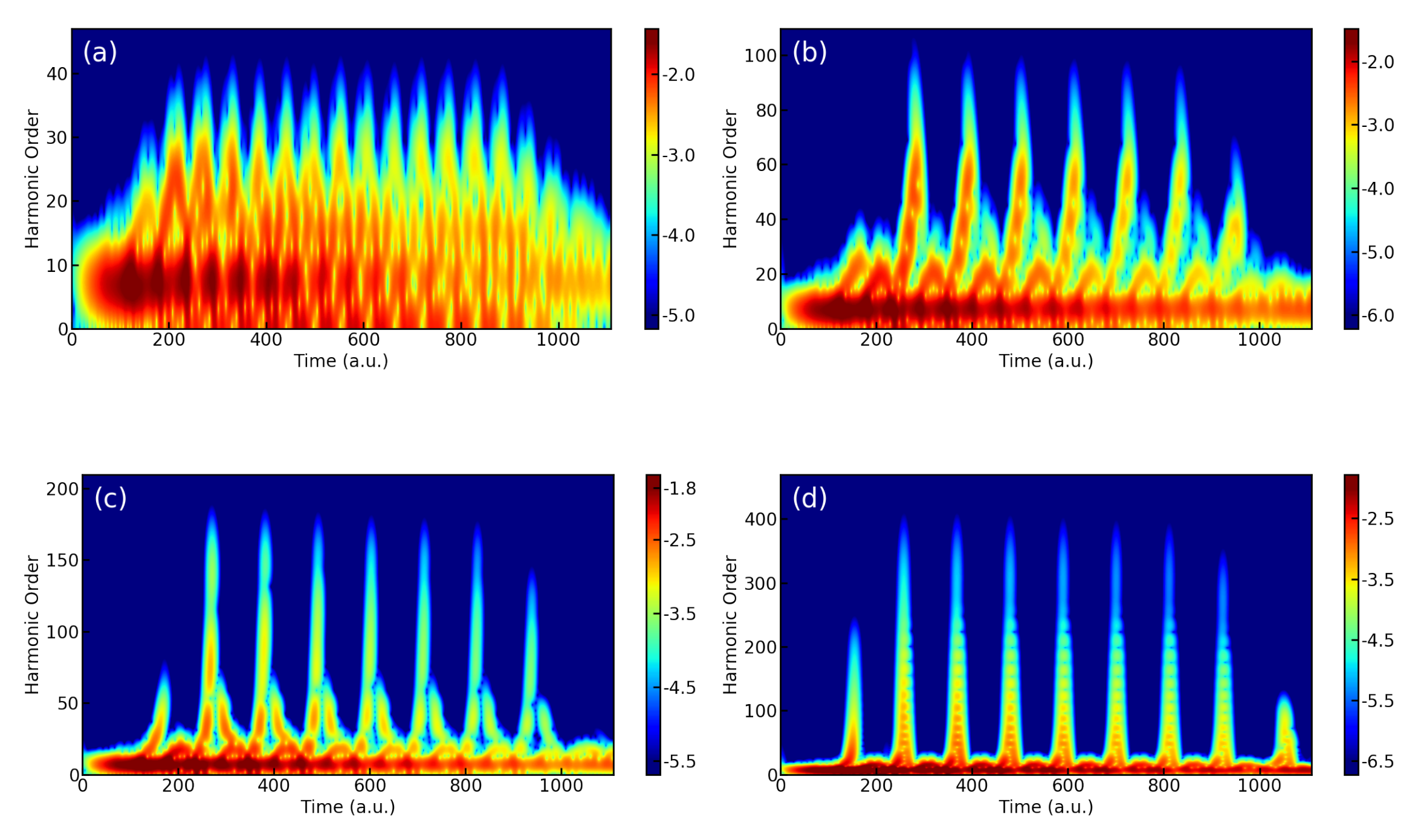}
\caption{(colour online). Same as figure~\ref{fig3} for $x_{\textrm{lim}} = \pm 4.5 \alpha_{0}$ except populations of the states.} \label{fig5}
\end{figure}

The time-frequency map of the harmonics for the combined pulses is shown in Fig.~\ref{fig5}. 
For homogeneous field, the time-frequency maps are similar  for both $x_{\textrm{lim}} = \pm 4.5 \alpha_{0}$ and $ \pm 7.5 \alpha_{0}$ (Figs.~\ref{fig5}(a) and \ref{fig3}(a)). 
When the strength of the inhomogeneity  is very small ($\beta = 0.01$),  the contribution from long-trajectory electrons is absent, as evident from Fig.~\ref{fig5}(b).  Additionally, since  the contribution from the long trajectories is weak, the harmonic spectrum does not change significantly after reducing the grid size. Furthermore, when  the strength of the inhomogeneity increases ($\beta = 0.02$ and 0.05), the contribution from high energetic short-trajectory electrons is missing along with the long-trajectory electrons.  This behaviour is expected because high energetic electrons travel farther away from the parent ion and their contributions are eliminated by reducing the grid size. 

 \begin{figure}[h]
\includegraphics[width= 17cm]{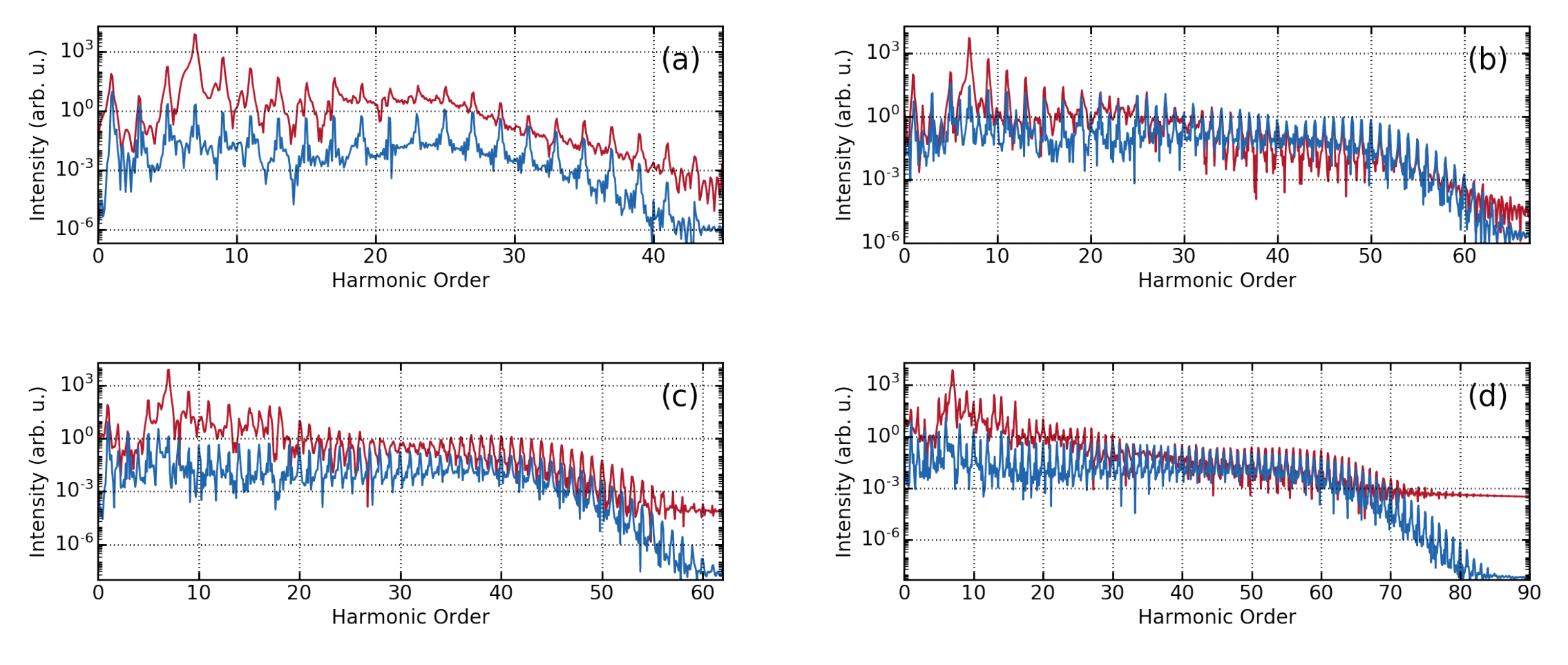}
\caption{(colour online). Same as figure~\ref{fig1} for $x_{\textrm{lim}} = \pm 1.5 \alpha_{0}$.} \label{fig6}
\end{figure}

 \begin{figure}[h]
\includegraphics[width= 17cm]{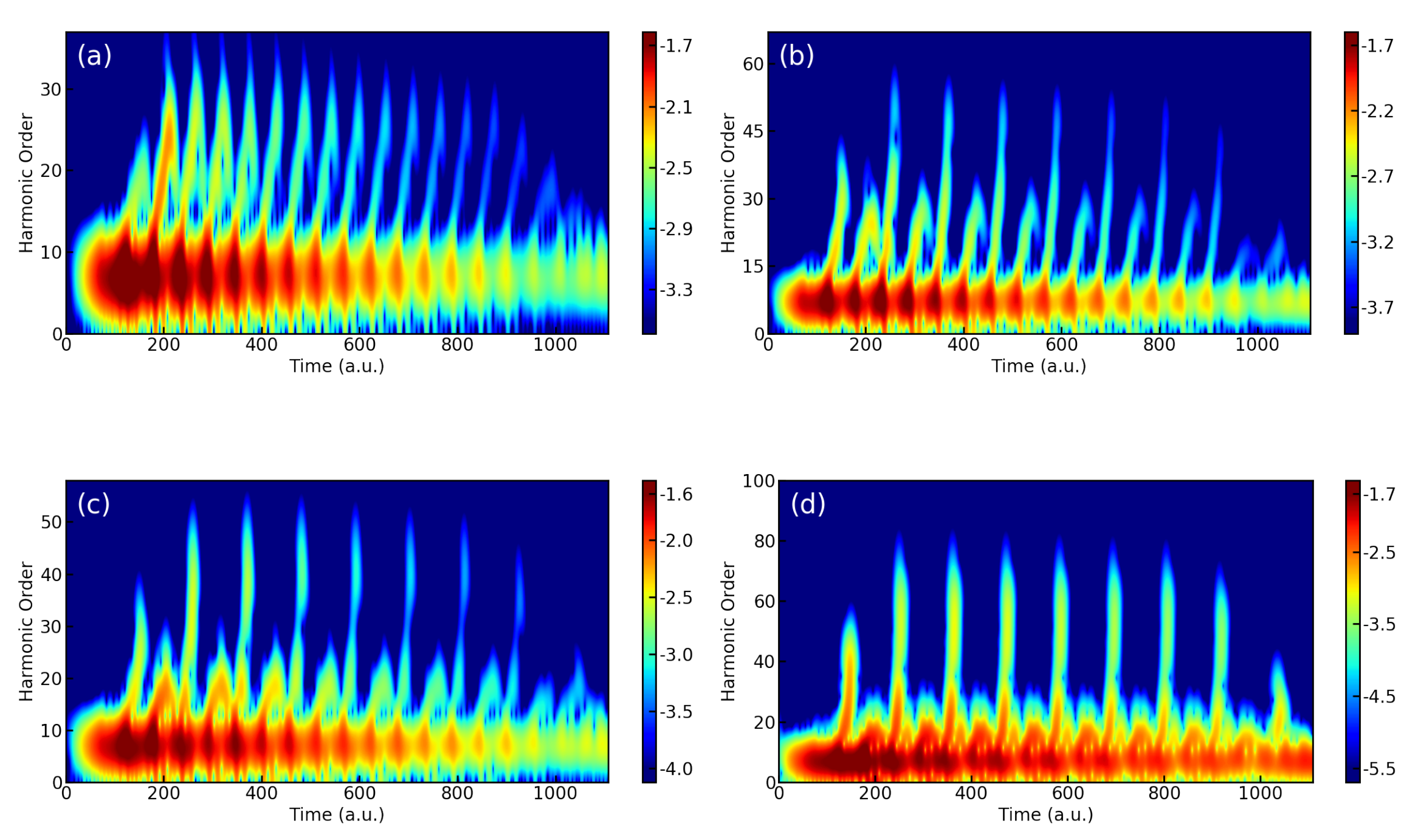}
\caption{(colour online). Same as figure~\ref{fig3} for $x_{\textrm{lim}} = \pm 1.5 \alpha_{0}$ except populations of the states.} \label{fig7}
\end{figure}

It is demonstrated that the oscillation in the HHG spectra is due to the interference between the electrons trajectories. 
The interference between the long and short trajectories occurs 
till the 70th and 225th harmonic orders for $\beta = 0.02$ and 0.05, respectively, as 
demonstrated in  Figs.~\ref{fig5}(c) and   \ref{fig5}(d). 
These findings are in agreement with the oscillations shown in Figs.~\ref{fig4}(c) and \ref{fig4}(d).

For small inhomogeneity ($\beta = 0.01$),  the harmonic spectrum do not change significantly by reducing the grid size to 
$\pm 4.5 \alpha_{0}$. However, the spectrum indeed changes drastically by further reducing the grid size to  $\pm 1.5 \alpha_{0}$
(Fig.~\ref{fig6}(b)). In turn, the cut-off is reduced for the dominant pairs of trajectories as well and the presence of the other pair of trajectories is visible from the Gabor profile (see Fig.~\ref{fig7}(b)).
For $\beta = 0.02$ and 0.05, the reduction in the energy cut-off is escalated further due to the even more stringent confinement of the electron motion.  The contribution from the short trajectories is further eliminated, as reflected from the time-frequency map in Figs.~\ref{fig7}(c) and~\ref{fig7}(d).  With the removal of long trajectories by confining the electron motion strongly, 
no interference now occurs, which is indicated by the absence of the oscillatory behaviour in the HHG (Fig.~\ref{fig6}).  
However, a slow oscillation  in the  intensity of the harmonics, generated by the combined seed and fundamental pulses, is present, see e.g. the red colour curves in Figs.~\ref{fig6}(b) and~\ref{fig6}(c).  By inspecting the corresponding time-frequency map carefully, a significant reduction in intensity is visible at the 40th and 30th harmonic order for $\beta = 0.01$ and 0.02, respectively. For $\beta = 0.01$ a similar intensity reduction is noticed at the 40th harmonic order in the map for other values of the grid size extension (Figs.~\ref{fig3}(b) and ~\ref{fig5}(b)).  
When the seed pulse is introduced, the strength of the generated harmonics decreases 
significantly during later cycles of the pulse as evident from the Gabor profiles. The reason behind this decrease is attributed to the  drastic depletion of the ground state population and this finding is true for all values of the inhomogeneity strengths and spatial grid sizes. 

In all scenarios,  the range of the recombination time for the  trajectories is reduced as the strength of the inhomogeneity increases, which is reflected  by the larger slope values of the trajectories. Larger slopes can be exploited to generate intense trains of attosecond pulses 
by choosing an appropriate window of frequencies.  The broadband energy spectrum  of the attosecond pulse train could be obtained by selecting wider frequency windows, e.g. , from the 100th to the 250th harmonic order in Fig.~\ref{fig3}(c).

\section{Conclusion}
In this work, the combined effect of plasmonic-enhanced fundamental and seed pulses is studied in HHG.  The efficiency of the HHG yield is enhanced by few-orders of magnitude as well as 
the HHG energy cutoff is also increased drastically. In the case of plasmonic-enhanced assisted HHG, the intensity of the harmonics is decreasing gradually with an in-build oscillation.  The time-frequency map and population dynamics of electronic states are used to 
understand the underlying mechanism of the generated spectra. 
The interplay of the population dynamics and the interference of short and long trajectories defines the oscillatory behaviour in the HHG yield. 
For higher values of spatial inhomogeneity, the harmonics are emitted in a short span of time 
and the emission times are well-separated. 
This feature enables the possibility to generate a train of intense attosecond pulses with a broadband spectrum. Plasmonic-enhanced fields, combined by seeded pulses, would pave the way for a more precise control of the HHG features.

\section*{Acknowledgments}
 
G. D. acknowledges the Ramanujan fellowship (SB/S2/ RJN-152/2015). Supported by the project Advanced research using high intensity laser produced photons and particles (CZ.02.1.01/0.0/0.0/16\_019/0000789) from European Regional Development Fund (ADONIS). We acknowledge the Spanish Ministry MINECO (National Plan 15 Grant: FISICATEAMO No. FIS2016-79508-P, SEVERO OCHOA No. SEV-2015-0522, FPI), European Social Fund, Fundaci\'o Cellex, Generalitat de Catalunya (AGAUR Grant No. 2017 SGR 1341 and CERCA/Program), ERC AdG OSYRIS, EU FETPRO QUIC, and the National Science Centre, Poland-Symfonia Grant No. 2016/20/W/ST4/00314. 
  

\end{document}